\documentclass[twocolumn,showpacs,amsmath,amssymb,prc]{revtex4}
\usepackage{mathrsfs}

\usepackage{graphicx,color}
\usepackage{dcolumn}
\usepackage{bm}

\begin{document}


\title{Systematic calculations of $\alpha$-decay half-lives with an improved empirical formula}

\author{Z. Y. Wang$^1$}
\author{Z. M. Niu$^1$}
\author{Q. Liu$^1$}\email{quanliu@ahu.edu.cn}
\author{J. Y. Guo$^1$}\email{jianyou@ahu.edu.cn}

\affiliation{$^1$School of Physics and Material Science, Anhui
University, Hefei 230039, China}

\date{\today}

\begin{abstract}
Based on the recent data in NUBASE2012, an improved empirical
formula for evaluating the $\alpha$-decay half-lives is presented,
in which the hindrance effect resulted from the change of the ground
state spins and parities of parent and daughter nuclei is included,
together with a new correction factor for nuclei near the shell
closures. The calculated $\alpha$-decay half-lives are found to be
in better agreements with the experimental data, and the
corresponding root-mean-square (rms) deviation is reduced to $0.433$
when the experimental $Q$-values are employed. Furthermore, the
$Q$-values derived from different nuclear mass models are used to
predict $\alpha$-decay half-lives with this improved formula. It is
found that the calculated half-lives are very sensitive to the
$Q$-values. Remarkably, when mass predictions are improved with the
radial basis function (RBF), the resulting rms deviations can be
significantly reduced. With the mass prediction from the latest
version of Weizs\"{a}cker-Skyrme (WS4) model, the rms deviation of
$\alpha$-decay half-lives with respect to the known data falls to
$0.697$.
\end{abstract}

 \maketitle


\section{Introduction}

$Alpha$-decay is a very important process in the field of nuclear
physics. It was first observed as an unknown radiation by Becquerel
in 1896 and further formulated empirically by Geiger and Nuttall in
1911~\cite{Geiger1911PM}. Afterwards, applying the theory of quantum
mechanics in the field of nuclear physics, Gamow~\cite{Gamow1928ZP}
and Condon and Gurney~\cite{Condon1928Nature} independently
described the spontaneous $\alpha$ decay as a quantum tunnelling
effect through the potential barrier leading from the parent nucleus
to the two emitted fragments: the $\alpha$ particle and the daughter
nucleus. As a powerful tool, $\alpha$ decay can be used to
investigate the low-energy structure of unstable nuclei, such as the
ground-state energy, the ground-state half-life, the shell effects,
and so on~\cite{Ren1987PRC, Horiuchi1991NPA, Firestone1996NewYork,
Lovas1998PR, Garcia2000JPG, Audi2003NPA, Gan2004EPJA,
Seweryniak2006PRC, Leppanen2007PRC}. On the experimental side, the
observation of $\alpha$-decay chains from unknown parent nuclei to
known nuclei has been a reliable method used to identify different
superheavy elements (SHEs) and isomeric states as
well~\cite{Hofmann2000RMP, Ginter2003PRC, Oganessian2005PRC,
Oganessian2012PRL}.

Up to now, on the basis of Gamow's theory, the absolute
$\alpha$-decay width has been estimated by many theoretical
calculations~\cite{Buck1992PRC, Xu2005NPA, Basu2003PLB,
Chowdhury2007PRC, Zhang2006PRC, Sharma2005PRC, Pei2007PRC}, which
employ various approaches, such as the cluster
model~\cite{Buck1992PRC, Xu2005NPA, Ni2009PRC, Ni2010PRC}, the
density dependent M3Y (DDM3Y) effective
interaction~\cite{Basu2003PLB, Chowdhury2007PRC}, the generalized
liquid drop model (GLDM)~\cite{Zhang2006PRC}. Different from the
cluster model, also some other theoretical models have been proposed
in the pursuit of a microscopic description of $\alpha$ decay, such
as the shell model and the fission-like model~\cite{Varga1992PRL,
Buck1993ADNDT, Poenaru1985PRC, Poenaru2011PRL}.

Moreover, many simple empirical formulas were also exploited to
analyze the $\alpha$ decay~\cite{Royer2000JPG, Viola1966JINC,
Sobiczewski1989PLB, Parkhomenko2005APPB, Denisov2009ADNDT,
Denisov2009PRC, Royer2008PRC, Royer2010NPA, Dong2010NPA,
Wang2010EPJA, Wang2014JPG, Ni2008PRC, Qian2011PRC, Ren2012PRC,
Qian2014PRC, Bao2014PRC}. Among these studies, Viola and Seaborg
proposed a semi-empirical formula to analyze $\alpha$-decay
half-lives for the heavy elements ($A\geq140$)~\cite{Viola1966JINC},
which has been often used up to the present day~\cite{Royer2010NPA,
Qian2011PRC}. In 2000, Royer developed other simple analytical
formula by fitting on a completed set of 373 $\alpha$
emitters~\cite{Royer2000JPG}, which can well reproduce the
experimental half-lives of the favored $\alpha$-decay. Based on the
above two works, Sobiczewski and
Parkhomenko~\cite{Parkhomenko2005APPB} also presented a simple
phenomenological formula for describing $\alpha$-decay half-lives of
heavy (above $^{208}$Pb) and superheavy nuclei. In previous works,
the ground state spins and parities of parent and daughter nuclei
were conventionally ignored. However, when the spin and parity
values of parent and daughter nuclei are different, then the emitted
$\alpha$ particle carries out nonzero angular momentum $l$. Because
of this, the orbital moment of emitted $\alpha$ particle should be
taken into account in an accurate approach for $\alpha$ decay. In
this way, Denisov and Khudenko~\cite{Denisov2009PRC} explored a
carefully updated and selected partial $\alpha$-decay half-life data
set of 344 ground-state-to-ground-state (g.s.-to-g.s.) $\alpha$
transitions, and presented sets of simple relations for evaluating
the half-lives of $\alpha$ transitions. After that,
Royer~\cite{Royer2010NPA} proposed other sets of analytical formulas
for $\log_{10} T_{1/2}$ depending or not on the angular momentum of
the $\alpha$ particle from an adjustment on the similar experimental
data set mentioned in Ref.~\cite{Denisov2009PRC}. Recently, Dong
$\emph{et al.}$~\cite{Dong2010NPA} extended the Royer's formula by
taking into account the contribution of the centrifugal barrier, and
then proposed a novel law, named the improved Royer's formula, for
calculating $\alpha$-decay half-lives, which was proved to work
well~\cite{Wang2010EPJA, Wang2014JPG}. Very recently, Ren $\emph{et
al.}$~\cite{Ren2012PRC} proposed a new Geiger-Nuttall law where the
effects of the quantum numbers of $\alpha$-core relative motion as
well as the possible effect of angular momentum and parity of
$\alpha$ particle are naturally embedded in the law.

In the empirical formulas, the alpha-decay half-life is determined
by the proton number $Z$, the number of nucleons in nucleus $A$, and
the value of decay energy $Q$, which can be derived from the nuclear
mass~\cite{Audi2003NPA}. Half-life calculations are very sensitive
to the choice of $Q$-values, so reliable theoretical predictions of
the nuclear mass, leading to precise $Q$-values, are essential to
study $\alpha$ transitions when these nuclear masses can not be
determined experimentally. In the past decades, many theoretical
calculations were performed to extrapolate nuclear masses. One
conventional method is the local mass relations, which have a high
precision of prediction for nearby nuclei, such as the Garvey-Kelson
relations~\cite{Garvey1969RMP}, residual proton-neutron
interactions~\cite{Zhang1989PLB, Fu2010PRC}, Coulomb-energy
displacement~\cite{Sun2011SCPMA, Kaneko2013PRL}, and systematics of
$\alpha$-decay energies~\cite{Dong2011PRL}. The other method relies
on global mass models, which are usually believed to have a better
ability of mass extrapolation for nuclei far from the known region,
such as the macroscopic-microscopic finite-range droplet model
(FRDM)~\cite{Moller1995ADNDT}, the Weizs\"{a}cker-Skyrme (WS)
model~\cite{Wang2014PLB}, the microscopic Hartree-Fock-Bogoliubov
(HFB) theory with a Skyrme force~\cite{Goriely2013PRC} and the
relativistic mean-field (RMF) model~\cite{Geng2005PTP}.

In this paper, we synthetically consider the centrifugal effect and
the hindrance of $\alpha$ emission with odd values of $l$, and then
renovate the improved Royer's formula by introducing two new
empirical terms. The effect of angular momentum and parity of
$\alpha$ particle is embedded remarkably, together with a new
correction factor for nuclei near the shell closures. Comparisons
about the root-mean-square (rms) deviation are performed, and the
numerical results show that our formula can well reproduce the
experimental half-lives of $341$ $\alpha$ transitions between the
ground states of nuclei. Moreover, we also consider different mass
models to provide the $Q$-values for $\alpha$-decay half-lives.
Especially, when the radial basis function (RBF) approach is
canonically introduced to improve the precision of the mass models.
A decrease of the rms deviations is obtained, which indicates that
the calculated $\alpha$-decay half-lives after embedding the RBF
approach agree better with the experimental data.

This paper is organized as follows. In section II, the theoretical
method of calculating $\alpha$-decay half-lives is briefly
described. In section III, the numerical results of $\alpha$-decay
half-lives as well as some detailed discussions are given, including
some comparisons between different empirical formulas and different
mass models with (without) the RBF approach. Finally, we render a
concise summary in section IV.

\section{The theoretical method}
\subsection{Input experimental data}

In this paper, a recent $\alpha$-decay data of $341$ nuclei is taken
from NUBASE2012~\cite{Audi2012CPC}, and then a fitting procedure on
this data set leads to an improved empirical formula. Moreover, the
following study is initially restricted to the g.s. to g.s. $\alpha$
transitions, which are known as partial $\alpha$ decay of the parent
nucleus. In view of this, it is necessary to take into account
experimental values of the branching ratio ($R$) for $\alpha$
transitions between the ground state of parent nuclei and various
states of daughter nuclei for correct extracting of the half-lives
for the g.s. to g.s. $\alpha$ decay.

According to the updated information of Audi $\emph{et
al.}$~\cite{Audi2012CPC}, there are totally about $700$ nuclei for
all possible $\alpha$ transitions with different $R$-values. The
nuclei with well-defined experimental $R$-values are adopted in this
paper. As a result, our data set contains $341$ nuclei with their
$R$-values being in the region of $1\%\leq R \leq 100\%$. These
nuclei can be divided into four categories: $123$ even $Z$-even $N$
(e-e), $93$ even $Z$-odd $N$ (e-o), $79$ odd $Z$-even $N$ (o-e), and
$46$ odd $Z$-odd $N$ (o-o) nuclei.

\subsection{Spin and parity selection rule}

In general, for e-e case, the spin and parity values of parent and
daughter nuclei are usually ignored. However, for the other three
cases, the transitions may occur with different spins and parities
of the parent and daughter nuclei and, consequently, the $\alpha$
particle may take away an nonzero angular momentum $l$. Accordingly,
the effect of the orbital moment of emitted $\alpha$ particle should
be taken into account in an accurate approach for $\alpha$ decay. We
select $l$ in accordance with the spin-parity selection rule,
similar as that in Refs.~\cite{Denisov2009ADNDT, Denisov2009PRC},
\begin{eqnarray}
l_{min}= \left\{
\begin{array}{l} \bigtriangleup_{j} \hskip 0.7 cm   \textrm{for} \hskip 0.1 cm \textrm{even} \hskip 0.1 cm
\bigtriangleup_{j} \hskip 0.1 cm \textrm{and} \hskip 0.1 cm \pi_{p} =\pi_{d},\\
\bigtriangleup_{j}+1 \hskip 0.1 cm   \textrm{for} \hskip 0.1 cm \textrm{even} \hskip
0.1 cm \bigtriangleup_{j} \hskip 0.1 cm \textrm{and} \hskip 0.1 cm \pi_{p}
\neq\pi_{d},\\
\bigtriangleup_{j} \hskip 0.7 cm   \textrm{for} \hskip 0.1 cm
\textrm{odd} \hskip 0.2 cm \bigtriangleup_{j} \hskip 0.1 cm
\textrm{and} \hskip 0.1 cm \pi_{p} \neq
\pi_{d},\\
\bigtriangleup_{j}+1 \hskip 0.1 cm   \textrm{for} \hskip 0.1 cm
\textrm{odd} \hskip 0.2 cm \bigtriangleup_{j} \hskip 0.1 cm
\textrm{and} \hskip 0.1 cm \pi_{p}
=\pi_{d},\\
\end{array} \right.
\end{eqnarray}
where $\bigtriangleup_{j}=|j_{p}-j_{d}|$, $j_{p}$, $\pi_{p}$,
$j_{d}$ and $\pi_{d}$ are spin and parity values of the parent and
daughter nuclei, respectively. It is worth noting that the orbital
angular momentum of the emitted $\alpha$ particle can have several
values according to the selection rule. However, for the sake of
simplicity, in the following calculations, the angular momentum $l$
of the emitted $\alpha$-particle is endowed with a minimum value
$l_{\textrm{min}}$. In addition, the values of spin and parity of
parent and daughter nuclei in this paper are taken from
Refs.~\cite{Denisov2009ADNDT, Goriely2013PRC, Audi2012CPC}.

\subsection{Analytical formula for the $\alpha$ decay half-lives}

In 2000, by fitting the experimental lifetimes of 373 emitters with
$R$-values close to 100\%, Royer gave an elementary analytical
formula~\cite{Royer2000JPG},
\begin{eqnarray}
   \log_{10}(T_{1/2})=a+bA^{1/6}\sqrt{Z}+\frac{cZ}{\sqrt{Q_\alpha}}.
\end{eqnarray}
Here $a$, $b$ and $c$ are the fitting parameters, $A$ and $Z$ are
the mass number and charge of parent nucleus, respectively,
$T_{1/2}$ is the half-life of $\alpha$ decay and given in seconds
while $Q$ the corresponding decay energy and given in MeV. This
formula can well reproduce the experimental half-lives of the
favored $\alpha$-decay. However, it dose not consider the angular
momentum in both unfavored and hindered $\alpha$ transitions. Just
because of this, the above formula was reconsidered by Dong
$\emph{et al.}$, and converted into a new version in terms of the
centrifugal contribution to the $\alpha$-nucleus
potential~\cite{Dong2010NPA},
\begin{eqnarray}
\log_{10}(T_{1/2})&=&a+bA^{1/6}\sqrt{Z}+\frac{cZ}{\sqrt{Q_\alpha}}\nonumber\\
&+&\frac{l(l+1)}{\sqrt{(A-4)(Z-2)A^{-2/3}}}.
\end{eqnarray}

In this paper, after detailed investigation about the explicit
dependence of the $\alpha$-decay half-lives, we propose a different
formula,
\begin{eqnarray}
\log_{10}(T_{1/2})&=&a+bA^{1/6}\sqrt{Z}+\frac{cZ}{\sqrt{Q_\alpha}}\nonumber\\
&+&\frac{d^{1-(-1)^l}l(l+1)}{\sqrt{(A-4)(Z-2)A^{-2/3}}}+S.
\end{eqnarray}
The four parameters $a$, $b$, $c$ and $d$ can be obtained by fitting
the new experimental data set, and then expressed in Tab.~\ref{tb1}.
It is interesting that we find the coefficient $d^{1-(-1)^l}$ is a
value less than $1$ for odd $l$ in both o-e and o-o cases, so this
may imply that the fourth term in the formula proposed in Eq. (3)
may overestimate the centrifugal effects for these two cases.

\begin{table}
\begin{center}
\caption{The parameters of our empirical formula found for e-e, e-o,
o-e, and o-o ranges of nuclei, respectively.} \label{tb1}
\begin{tabular}{lccccccccccccccc}
\hline\hline & $a$  & $b$  & $c$  & $d$  \\ \hline
e-e & -25.432 & -1.146 & 1.577 & ---    \\
e-o & -26.591 & -1.171 & 1.639 & 1.123  \\
o-e & -27.747 & -1.093 & 1.620 & 0.829  \\
o-o & -28.460 & -0.984 & 1.573 & 0.970  \\
\hline\hline
\end{tabular}
\end{center}
\end{table}

In Eq. (4), the first three terms are the same as those in
Refs.~\cite{Royer2000JPG, Dong2010NPA}, the fourth term is similar
to the corresponding one in Ref.~\cite{Dong2010NPA}, which can be
attributed to the distinct contribution of the centrifugal potential
$\frac{\hbar^2}{2\mu}l(l+1)$ to the total $\alpha$-nucleus potential
at small distances between daughter nucleus and $\alpha$ particle
while $l\neq0$~\cite{Denisov2009ADNDT}. The difference is that we
have introduced an additional factor $d^{1-(-1)^l}$ into this term
to account for the hindrance of the transition with the change of
parity. It is such an exponential form that the formula can well
reflect the changes of $alpha$-decay half-lives originated from $l$.
Due to the rearrangement of single-particle orbits and of the
nuclear spin orientation, the structure hindrance is relatively
weak~\cite{Karamian2007PRC} and, consequently, will not be discussed
here. The last empirical term is a phenomenological correction
factor and defined as: $S=0.5$ for $49\leq Z\leq 51$, $81\leq Z\leq
83$, $49\leq N\leq 51$, $81\leq N\leq 83$ or $125\leq N\leq 127$.
$S$ is introduced to mock up the systematic deviations of calculated
$\alpha$-decay half-lives with respect to the experimental data near
shell closures.

In addition, the rms deviation of the decimal logarithm of the
$\alpha$-decay half-life, in this paper, is defined as
\begin{eqnarray}
\sigma=\sqrt{\frac{1}{N}\sum^{N}_{i=1}[\log_{10}(T^{\textrm{theo.}}_{1/2
i})-\log_{10}(T^{\textrm{expt.}}_{1/2 i})]^2}.
\end{eqnarray}
$N$ is the number of nuclei used for evaluation of the rms
deviation. The $Q$-value for the $\alpha$-decay half-life can be
calculated using experimental mass data as $Q= M_{p}-(M_{d}+
M_{\alpha})$, $ M_{p}$, $ M_{d}$, and $ M_{\alpha}$ are the masses
of parent and daughter nuclei and $\alpha$ particle, respectively.

\section{Result and discussion}

\begin{figure}[h]
  \includegraphics[width=7cm]{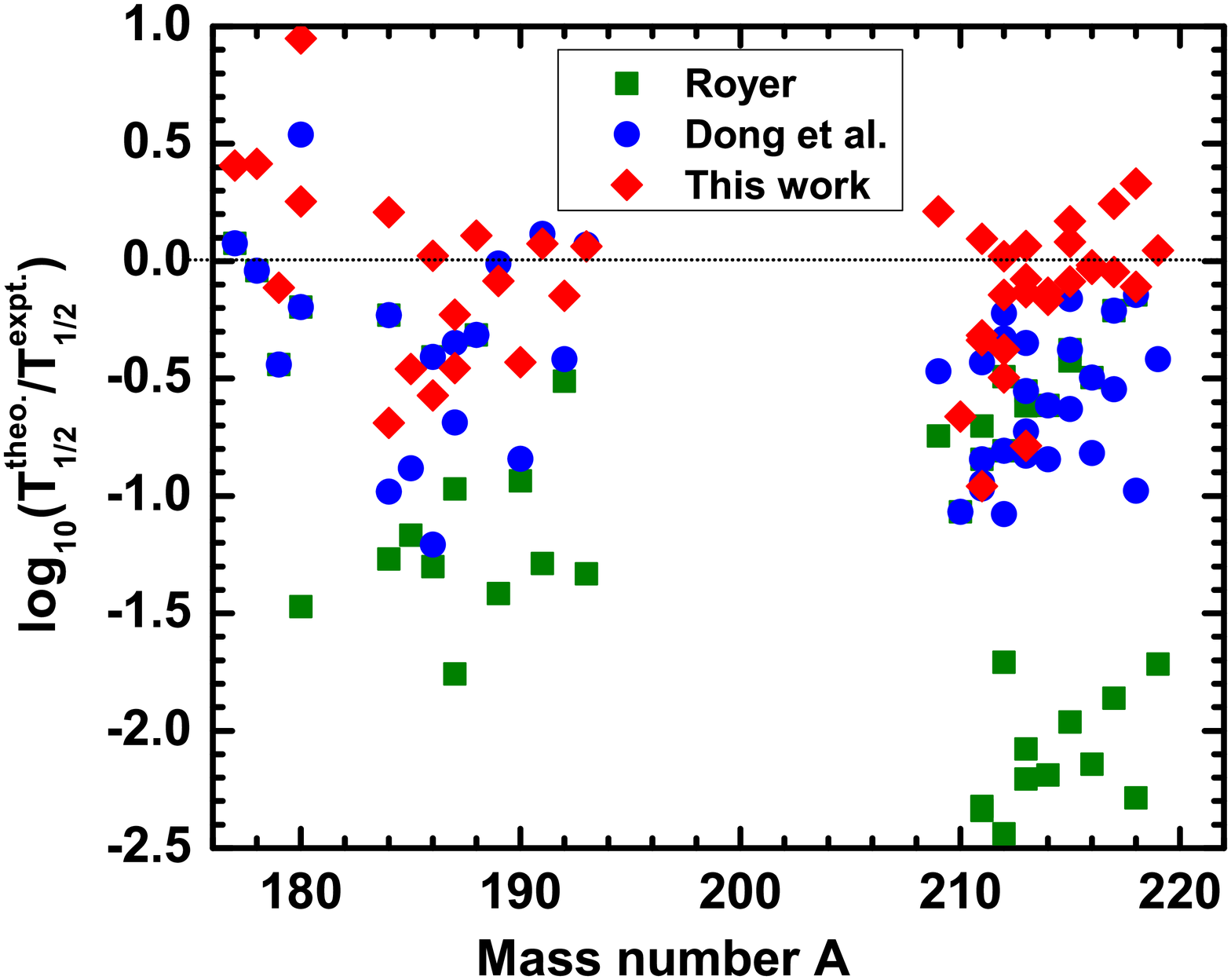}\\
  \caption{ (Color online) Logarithms of the ratios between theoretical
$\alpha$-decay half-lives calculated with different formulas and
experimental ones versus the mass number $A$ of the parent nucleus.
The red diamonds denote the results of this work, the blue circles
and the olive squares correspond to the results calculated with two
formulas proposed by Dong \emph{et al.} and Royer,
respectively.}\label{fig1}
\end{figure}

As proposed above, Royer's formula works well in the favored
$\alpha$-decay~\cite{Royer2000JPG}. However, when it is used to
evaluate the half-lives for nuclei with $Z$ and $N$ crossing the
$Z=82$ or $N=126$ shell closures, respectively, a dramatic large
deviation can be observed between calculated half-lives and
experimental ones, which is drawn clearly in Fig.~\ref{fig1} with
the squares corresponding to the results of Royer's formula in
Eq.(2). It can be seen in Fig.~\ref{fig1} that all squares are far
away from the dotted line, the calculated values of $\alpha$-decay
half-lives are always less than the experimental data for both
$Z=81-83$ isotopes and $N=125-127$ isotones. There are about a half
of the whole isotopes and isotones with the ratios between
calculated values and experimental ones being beyond a factor of ten
($\log_{10}10=1$). The systematic deviation is very apparent.

Then after including the centrifugal effect, Royer's formula was
extended to the unfavored $\alpha$-decay with a new form proposed in
Eq.(3)~\cite{Dong2010NPA}. With this formula, the same calculations
are performed on both the isotopic chain of $Z=81-83$ and the
isotonic chain of $N=125-127$, and we plotted the corresponding
results with the filled circles in Fig.~\ref{fig1}. It is obvious
that the values of
$\log_{10}(T^{\textrm{theo.}}_{1/2}/T^{\textrm{expt.}}_{1/2})$ for
the most nuclei (denoted by filled circles) are in the range from
$-1$ to $0$ and the whole points land close to the dotted line. It
means the calculated $\alpha$-decay half-lives with Eq. (3) agree
better than those deduced from Eq. (2) with the experimental data.
However, the deviation is still systematic with the mean absolute
value of
$\log_{10}(T^{\textrm{theo.}}_{1/2}/T^{\textrm{expt.}}_{1/2})$ being
about $0.5$. Accordingly, the systematic behavior of deviation is
decreased to some extend but not completely redressed. Then how to
overcome the deviation error between the calculated half-lives and
the experimental data?

As we know, accurate calculations of $\alpha$ transitions should
take into account the spins and parities of parent and daughter
nuclei and the angular momentum of the emitted $\alpha$ particle.
Particularly for the transition with large change of spin and with
the change of parity, the factor $d^{1-(-1)^l}$ can well simulate
the hindrance of $\alpha$ emission. Moreover, when $Z$ ($N$) goes
across the shell closure at $Z=82$ ($N=126$), the effect of the
closed shell results in a decrease of the $\alpha$-preformation
factor, which has been shown in the experimental
analysis~\cite{Hodgson2003PR}, that may be the reason of the
existence of the large deviations proposed above. In this way, we
introduce two new empirical terms to remedy the systematic
deviation, and a new analytical formula is proposed in Eq. (4).
Applying our formula to both $Z=81-83$ isotopes and $N=125-127$
isotones leads to a desired effect, which can be seen in
Fig.~\ref{fig1} with the red points. One can see that most of the
red points float around the dotted line, the absolute values of
$\log_{10}(T^{\textrm{theo.}}_{1/2}/T^{\textrm{expt.}}_{1/2})$ are
generally less than $0.4$ for almost all of the parent nucleus.
Accordingly, the systematic behavior of deviation is decreased to a
large extent.

\begin{figure}[h]
  \includegraphics[width=7.5cm]{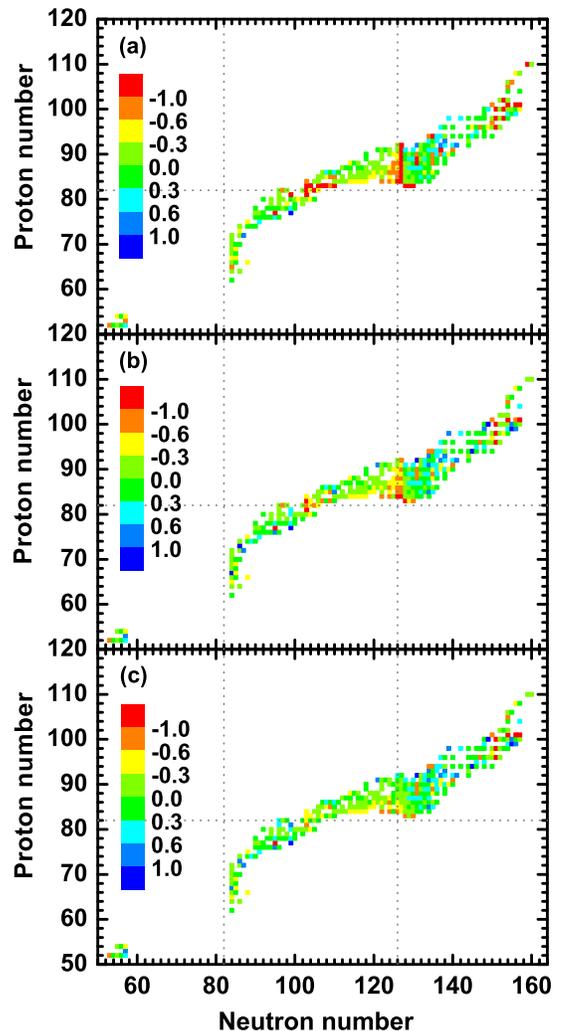}\\
  \caption{ (Color online) Decimal logarithms of $T^\textrm{{theo.}}_{1/2}/T^\textrm{{expt.}}_{1/2}$ for 341
nuclei derived from the different formulas, for which calculations
are performed with the present parameter set of this work. Panel (a)
denotes the case with Royer's formula proposed in Eq. (2), panel (b)
corresponds to the case with the improved Royer's formula proposed
in Eq. (3), and panel (c) is designated the result obtained with our
formula proposed in Eq. (4), respectively. Dotted lines denote the
magic numbers. }\label{fig2}
\end{figure}

For further insight of the dependence on $l$ and $S$, a comparison
between the experimental $\alpha$-decay half-lives and the
calculated ones for whole $341$ nuclei with the three formulas
proposed above is performed, and is drawn in Fig.~\ref{fig2}, in
which the results of the empirical formulas are calculated with the
parameter values from the present work. In Fig.~\ref{fig2}, Panel
(a) denotes the decimal logarithm of
$T^\textrm{{theo.}}_{1/2}/T^\textrm{{expt.}}_{1/2}$ for $341$ nuclei
with Royer's formula proposed in Eq. (2). One can see that around
the shell closures ($Z=82$ and $N=126$), the calculated half-lives
are much shorter than the experimental ones, the ratios between
calculated values and experimental ones are beyond a factor of ten
($\log_{10}(T^{\textrm{theo.}}_{1/2}/T^{\textrm{expt.}}_{1/2})=-1$).
The rms deviation for the full data set is $0.659$. In panel (b),
the formula proposed in Eq. (3) is applied. It is obvious that the
agreements with the experimental data are visible around the shell
closures. The amounts of discrepancy also decrease in other fields,
and the rms deviation is reduced to $0.487$, all of which indicate
the extending of Royer's formula by considering the centrifugal
effect is successful. Panel (c) is designated the result obtained
with our formula proposed in Eq. (4). In this panel, it is
worthwhile to note that systematic behavior of deviation is rapidly
decreased. The absolute values of
$\log_{10}(T^{\textrm{theo.}}_{1/2}/T^{\textrm{expt.}}_{1/2})$ is
optimized in many fields, and averagely about $0.27$ for nuclei near
the shell closures. Our rms deviation is only $0.433$. Hence, the
introduction of the term $S$ and parity correction $d^{1-(-1)^l}$
improves strongly the efficiency of the formula proposed in Eq. (4),
even though they are semi-empirical. By the way, at the top
right-hand corner in Panel (a)-(c), the ratios between calculated
values and experimental ones are beyond a factor of ten for some
isotopes or isotones and not decreased after introducing $l$ and
$S$. The reason may be that the SHE experiments are very difficult
and usually few decay events are observed. Hence the experimental
error bar is relatively large in the measurement of both decay
energies and half-lives.

\begin{table}
\begin{center}
\caption{Comparisons between the rms deviations of the experimental
and the calculated $\alpha$-decay half-lives for the full data set
as well as for e-e, e-o, o-e, and o-o subsets, respectively. The
first column contains the symbols for corresponding formulas.}
\label{tb2}
\begin{tabular}{lccccccccccccccc}
\hline\hline & Total  & e-e  & e-o  & o-e  &  o-o  \\
  &  ($N=341$) & ($N=123$) & ($N=93$) & ($N=79$) & ($N=46$) \\ \hline
Eq. (2)   & 0.587 & 0.298 & 0.713 & 0.610 & 0.810   \\
Eq. (3) & 0.481 & 0.298 & 0.574 & 0.486 & 0.637 \\
Eq. (4) & 0.433 & 0.267 & 0.521 & 0.434 & 0.574  \\
Ref.~\cite{Denisov2009PRC} & 0.536 & 0.350 & 0.489 & 0.539 & 0.689   \\
Ref.~\cite{Royer2010NPA} & 0.561 & 0.285 & 0.509 & 0.548 & 0.747 \\
\hline\hline
\end{tabular}
\end{center}
\end{table}

The detailed numerical results for the full data set are listed in
Tab.~\ref{tb2} as well as for e-e, e-o, o-e, and o-o subsets,
respectively. Note that in order to study the effect of $S$ and $l$
on the accuracy of the formula, the parameters used in either
Royer's formula~\cite{Royer2000JPG} or the improved Royer's
formula~\cite{Dong2010NPA} have been recalculated by fitting the new
experimental data set. After this, it is found that the
corresponding rms deviations deduced from the above two formulas are
$0.587$ and $0.481$, less than the old $\sigma$ of $0.646$ and
$0.504$ with the parameters taken from Refs.~\cite{Royer2000JPG,
Dong2010NPA}, respectively. It means that the accuracies of these
two formulas are improved apparently, but still lower than that of
our formula with the standard $\sigma$-values being only $0.433$.
Moreover, a further comparison is performed between our formula with
the well-known empirical formulas presented by Denisov and
Khudenko~\cite{Denisov2009PRC} and Royer in
2010~\cite{Royer2010NPA}, respectively. The calculations are also
performed with the new parameters determined by fitting the updated
experimental data set, and then lead to the corresponding
$\sigma$-values of $0.536$ and $0.561$, respectively. The detailed
rms deviations for both full data set and four subsets are displayed
in Tab.~\ref{tb2} in the fourth and fifth lines, from which a
similar conclusion is attained in comparison with our formula. That
is our analytical formula has the smallest values of the rms
deviations for both full data set and most of the subsets except for
the e-o case, in which the rms deviations resulted from
Refs.~\cite{Denisov2009PRC, Royer2010NPA} are smaller. All these
indicate a higher accuracy in our formula than either one in
Ref.~\cite{Denisov2009PRC} and Ref.~\cite{Royer2010NPA}.

\begin{figure}[h]
  \includegraphics[width=8.5cm]{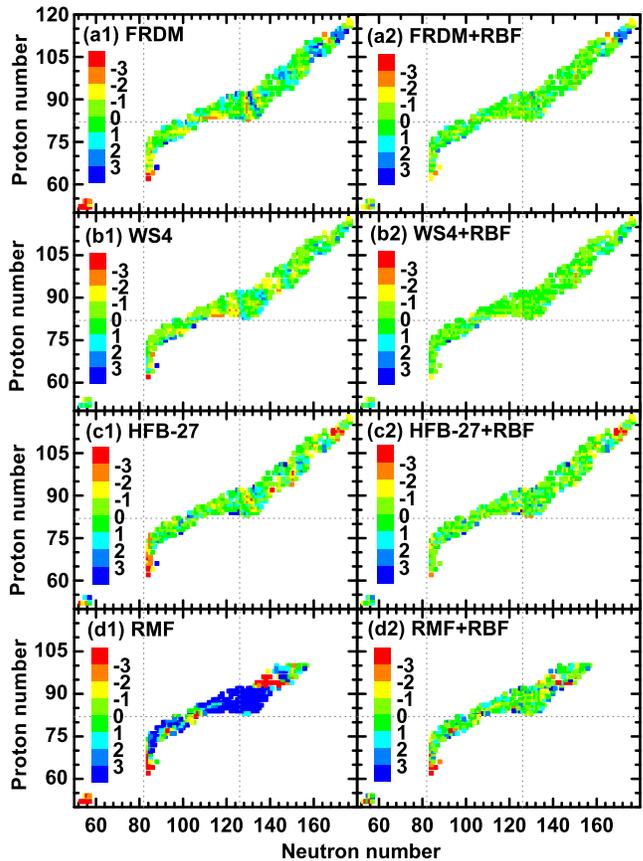}\\
  \caption{(Color online) The same as Fig.~\ref{fig2}, but for
different four mass models without (with) the RBF approach. The left
four panels (a1-d1) correspond to the case without the RBF approach,
while the right ones to the case with the RBF approach.
}\label{fig3}
\end{figure}

\begin{table}
\begin{center}
\caption{The same as Tab.~\ref{tb2}, but for different calculated
mass tables.} \label{tb3}
\begin{tabular}{lccccccccccccccc}
\hline\hline &  Model  & Model+RBF  \\\hline
FRDM & 1.622 &  0.968 \\
WS4 & 1.163 & 0.697 \\
HFB27 & 1.617 &  1.128\\
RMF & 5.822 &  1.511\\
Bhagwat & 1.184 & 0.862\\
DZ10 & 1.644 &  0.979\\
DZ31 & 1.597 & 1.035\\
KUTY & 1.184 & 0.890\\
\hline\hline
\end{tabular}
\end{center}
\end{table}

To examine the predictive ability of nuclear mass model on
$\alpha$-decay half-life, the improved formula expressed in Eq. (4)
is employed to calculate $\alpha$-decay half-lives with the
theoretical $Q$-values. In this work, eight nuclear mass models are
taken into account, i.e., the FRDM model~\cite{Moller1995ADNDT}, the
latest version of WS (WS4) model~\cite{Wang2014PLB}, the recent
version of the HFB model (HFB-27)~\cite{Goriely2013PRC}, the RMF
model with TMA effective interaction~\cite{Geng2005PTP}, the
Koura-Tachibana-Uno-Yamada (KTUY) model~\cite{Koura2005PTP}, the
Duflo-Zuker formulas~\cite{Dufloand1995PRC, Zuker2008PMFS} and the
Bhagwat formula ~\cite{Bhagwat2014PRC}. For simplicity, the
deviations of
$\log_{10}(T^{\textrm{theo.}}_{1/2}/T^{\textrm{expt.}}_{1/2})$ for
four mass models are drawn, and without loss of generality, FRDM,
WS4, HFB-27 and RMF are taken as examples (see Fig.~\ref{fig3}). The
corresponding numerical results of rms errors for eight mass models
are listed in Tab.~\ref{tb3}. As can be seen in Tab.~\ref{tb3}, the
descriptions of $\alpha$-decay half-lives for these mass models are
generally within the same order of magnitude except for the RMF
model. The smallest rms deviation of $1.163$ is obtained from the
WS4 model, which indicates the WS4 model has the best accuracy of
$Q_\alpha$ prediction among mass models considered here. However,
the RMF model significantly overestimates the $\alpha$-decay
half-lives for nuclei around $N=126$ and in some other fields (see
panel (d1) in Fig.~\ref{fig3}), so it is necessary to improve the
mass precision of RMF model for reliably predicting $\alpha$-decay
half-lives.

An efficient method to enhance the predictive power of mass models
is the radial basis function (RBF) approach, which was first
introduced into this field by Wang $\emph{et
al.}$~\cite{Wang2011PRC}. The basic formulas of the RBF approach
have been detailed in our previous works~\cite{Niu2013PRC,
Zheng2014PRC}. So we will not repeat it anymore hereafter. For
convenience, the mass model improved by the RBF approach is denoted
with the Model+RBF henceforth, e.g. FRDM+RBF and RMF+RBF. After
employing the RBF approach, the new $\alpha$-decay half-lives can be
deduced with the optimized $Q$ values derived by the above mass
equation, and subsequent comparisons between the new $\alpha$-decay
half-lives and the experimental data are performed and exhibited in
Tab.~\ref{tb3}. One can see from Tab.~\ref{tb3} that the systematic
rms deviations decrease remarkably after using the RBF approach for
all the eight mass models. Taking RMF model for an instance, the rms
deviation is $5.822$ before applying the RBF approach, and then
significantly decreases to $1.511$ with refinement by the RBF
approach. From Fig.~\ref{fig3}, one can also see that the calculated
$\alpha$-decay half-lives with RBF approach result in better
agreements with the experimental ones for RMF model than that
without RBF approach. The effect of systematic correction is
impressively clear. Meanwhile, there are the common cases for the
other mass models, all of which provide strong support for the
reliability of the RBF approach and its usefulness for eliminating
the discrepancy.

\section{Summary}

In summary, by fitting the new experimental data of $341$ nuclei, we
have proposed an empirical formula for systematically calculating
the half-lives of $\alpha$ transitions between the ground states of
parent and daughter nuclei. Within this new formula, we consider the
hindrance effect resulted from the changes of spin and parity of
parent and daughter nuclei and a phenomenological correction factor
for the nuclei near the shell closure. The calculations of
$\alpha$-decay half-lives with some other well-known empirical
formulas are also performed. By comparing, it is found that our
formula results in the smallest values of the rms deviation for both
full data set and any subset. In other words, the precision in our
formula is higher than that in the previous methods. Furthermore, we
apply the improved formula to eight calculated mass tables. Using
the $Q$-values resulted from these mass tables, the subsequent
half-life outcomes indicate that WS4 model has the best accuracy of
$Q_\alpha$ prediction. Then in order to improve the predictive
ability of the mass model, the RBF approach is adopted. As a result,
a rapid decrease of the rms deviation is obtained, which indicates
the calculated $\alpha$-decay half-lives, after embedding the RBF
approach, are in better agreements with the experimental data.\\

\section{Acknowledgements}

We thank Jiangming Yao and Zhipan Li for helpful discussions. This
work was partly supported by the National Natural Science Foundation
of China (Grants No. 11205004 and No. 11305002), the 211 Project of
Anhui University under Grant No. J01001319-J10113190081.




\end{document}